\documentclass[a4paper]{article}

\usepackage{INTERSPEECH2021}
\usepackage{cite}
\usepackage{caption}
\usepackage{subcaption}

\title{N-Singer: A Non-Autoregressive Korean Singing Voice Synthesis System for Pronunciation Enhancement}
\name{Gyeong-Hoon Lee, Tae-Woo Kim, Hanbin Bae, Min-Ji Lee, Young-Ik Kim, Hoon-Young Cho}
\address{Speech AI Lab., AI Center, NCSOFT, Republic of Korea}
\email{\{ghlee0304, ktw0114, bhb0722, leeminxji, youngik, hycho\}@ncsoft.com}

\begin{document}
%
\maketitle

\begin{abstract}
Recently, end-to-end Korean singing voice systems have been designed to generate realistic singing voices. However, these systems still suffer from a lack of robustness in terms of pronunciation accuracy. In this paper, we propose N-Singer, a non-autoregressive Korean singing voice system, to synthesize accurate and pronounced Korean singing voices in parallel. N-Singer consists of a Transformer-based mel-generator, a convolutional network-based postnet, and voicing-aware discriminators. It can contribute in the following ways. First, for accurate pronunciation, N-Singer separately models linguistic and pitch information without other acoustic features. Second, to achieve improved mel-spectrograms, N-Singer uses a combination of Transformer-based modules and convolutional network-based modules. Third, in adversarial training, voicing-aware conditional discriminators are used to capture the harmonic features of voiced segments and noise components of unvoiced segments. The experimental results prove that N-Singer can synthesize a natural singing voice in parallel with a more accurate pronunciation than the baseline model.
\end{abstract}


\section{Introduction}
\label{sec:intro}

The singing voice synthesis (SVS) system is a generative model that synthesizes singing voice from text and musical information such as lyrics, beat, note pitch and length. Recently, Korean SVS systems \cite{hankorean, atk, begansing} have attracted significant attention with the increase of Korean pop music (K-pop) popularity. To minimize the difference between real and synthesized singing voices, SVS systems should consider that synthesized singing voice must have not only accurate pronunciation, but also musical factors of the corresponding musical score. However, it is difficult to modeling linguistic and musical features simultaneously because singing voices generally have a longer vowel duration and a wider range of pitch than speech \cite{hifisinger}.

Previous deep-learning-based SVS systems adopted the architecture of autoregressive speech synthesis systems \cite{tacotron1, tacotron2, dctts} to synthesize a more realistic singing voice. Autoregressive neural SVS systems are designed to model text and musical information using auxiliary modules, such as pitch encoder \cite{atk} and duration module \cite{bytesing}, or feature composing method between linguistic and musical features \cite{hankorean}. Although autoregressive SVS systems outperform conventional parametric models \cite{unit, hmm, parametric}, these systems lack robustness; i.e., there are word skipping and repetition problems in synthesized singing voices because of incorrect attention alignment.

The adversarially trained end-to-end Korean SVS system (ATK) \cite{atk} has shown the best performance among Korean SVS systems \cite{hankorean, atk, begansing}. ATK is an autoregressive SVS system based on convolution \cite{dctts}. To generate a more accurate and pronounced singing voice, the mel-generator of ATK is designed to independently model the pronunciation information from text in an unsupervised manner using the phonetic enhancement mask decoder. In addition, ATK uses a conditional generative adversarial network \cite{gan} with a projection discriminator \cite{projection} to achieve cleaner linear spectrograms. However, it was observed that the ATK's synthesized singing voice pronunciation was inaccurate, and short phonemes were often skipped because convolution, which can only capture local features, caused incorrect attention alignment. Additionally, because ATK predicts the linear spectrogram, the complexity of the generator is too high, and the synthesis speed is slow.

To overcome the limitations of autoregressive generative models, non-autoregressive SVS systems have been proposed to synthesize singing in parallel \cite{seq2seqTransformer, xiaoicesing, xiaoicesingmulti, hifisinger}. In the non-autoregressive SVS systems, acoustic models can synthesize acoustic features such as mel-spectrogram faster and in a more robust manner than autoregressive systems using a Transformer-based on self-attention network, which has the advantage of capturing long-term dependencies in the input sequence \cite{attentionisallyouneed, fastspeech}. A length regulator and duration predictor were adopted so that each phoneme sequence matched the time length of the mel-spectrogram. However, in our preliminary experiments on Korean singing, we found that the pronunciation of singing voice generated by a previous SVS model is not accurate, given a pitch sequence with dynamic changes. 

In this paper, we propose N-Singer, a non-autoregressive Korean SVS system for synthesizing singing voices with more accurate pronunciation. Our contributions are listed below. First, for a singing voice that articulates pronunciation better, we independently model the linguistic features and musical features in mel-spectrograms by using two feed-forward Transformer (FFT)-based encoder-decoder modules. Second, to improve the quality of the synthesized mel-spectrograms, we jointly train the convolutional neural network (CNN)-based postnet and Transformer-based generator. Third, we use the GAN framework with voicing-aware conditional discriminators using pitch, voiced and unvoiced (V/UV) flag to efficiently train the acoustic model. Two discriminators deal with each voiced and unvoiced segment of a singing voice, effectively improving the performance of the generator. N-Singer uses Parallel WaveGAN (PWG) \cite{pwg} as a vocoder to synthesize a singing voice waveform in parallel. Experimental results show that N-Singer can synthesize mel-spectrograms with higher pronunciation accuracy and better sound quality, and is 36.7 times faster than the ATK on an NVIDIA V100 GPU.

\begin{figure*}[t]
  \centering
  \includegraphics[width=\linewidth]{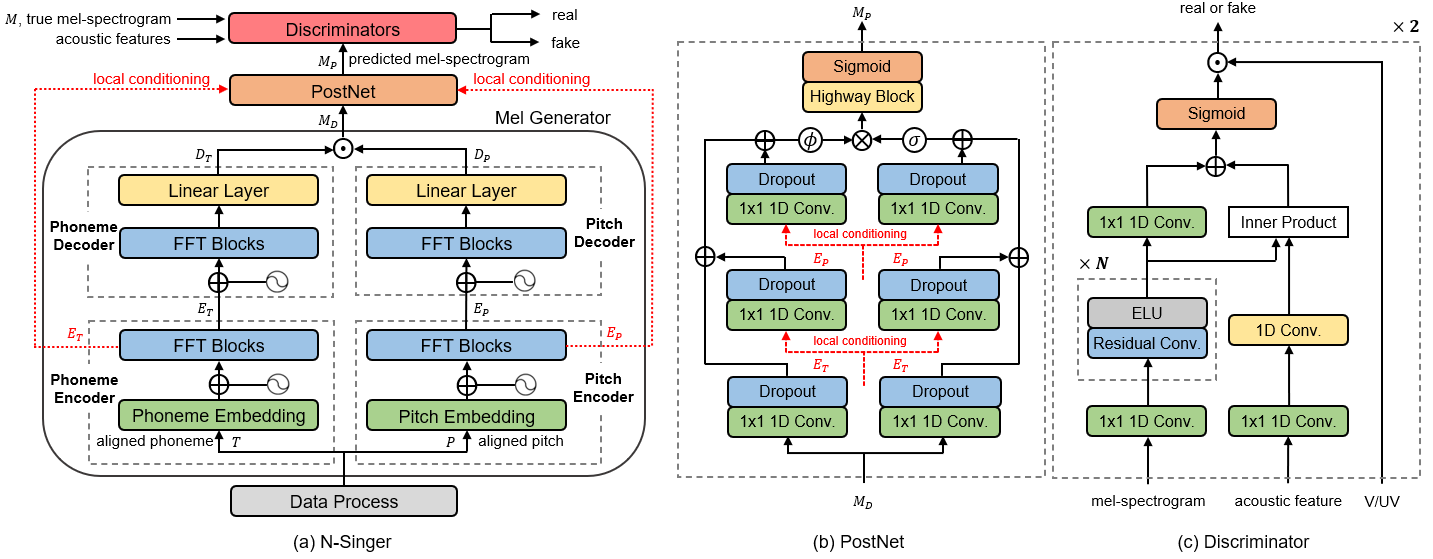}
  \caption{(a) is a whole structure of N-Singer, (b) is a postnet of N-Singer, and (c) is a voicing-aware discriminator of N-Singer}
  \label{fig:structure}
  \vspace{-0.3cm}
\end{figure*}

\section{Proposed Method}
\label{sec:proposed}

Our proposed Korean SVS system, N-Singer, consists of a mel-generator and postnet as shown in Figure 1. The mel-generator is composed of two encoder–decoder modules based on a Transformer, and each module uses lyrics and note pitch information independently. The CNN-based postnet generates an enhanced mel-spectrogram from a coarse mel-spectrogram using phoneme and pitch encoder outputs as local conditions.

\subsection{Input representation}

We converted the grapheme sequences in Korean lyrics to phoneme sequences using the Korean grapheme-to-phoneme algorithm. We rearranged the pitch sequence to manually match one Korean syllable with a single note pitch. Next, we expanded the phoneme and pitch sequences to be the same time length as the mel-spectrogram $M\in\mathbb{R}^{N \times L}$ by using the note duration in the musical score, where $N$ is the number of bins and $L$ is the number of frames of the mel-spectrogram. Generally, Korean syllables are composed of three components: onset, nucleus, and coda. In the aligned phoneme sequence $T \in \mathbb{R}^{1 \times L}$, we assigned onset and coda to a maximum of three frames and assigned the remaining frames to the nucleus because singing voices have a long vowel duration and the nucleus of a Korean syllable corresponds to the vowel. We converted each note pitch in the musical score into a pitch ID according to the standard musical instrument digital interface (MIDI) standard \cite{midi}. In the aligned pitch sequence $P\in\mathbb{R}^{1 \times L}$, the pitch corresponding to each syllable was repeatedly assigned to the frames to fit the syllable length. We referred to \cite{atk} for this method.

\subsection{Mel-generator}

The N-Singer's mel-generator consists of two encoder–decoder modules based on FastSpeech \cite{fastspeech}. In the encoder, the input sequence is embedded into a dense vector, added with positional encoding, and then passed through the FFT blocks. The FFT block consists of a self-attention network with multi-head attention and a 1D convolutional network \cite{fastspeech, xiaoicesing}. In the decoder, positional encoding is added to the encoded sequence and passed through the FFT block, passing through a linear layer with sigmoid activation. During training, the output $D_T \in \mathbb{R}^{N \times L}$ of the phoneme encoder-decoder contained the linguistic features related to pronunciation from the phoneme sequence $T \in \mathbb{R}^{1 \times L}$, while the pitch encoder–decoder's output $D_P \in \mathbb{R}^{N \times L}$ contained features related to F0 and harmonics from the pitch sequence $P \in \mathbb{R}^{1 \times L}$. Because the mel-generator can model pronunciation and harmonics separately, the phoneme encoder–decoder is not distorted by any input pitch sequence and can generate a mel-spectrogram containing more accurate pronunciation information. The output mel-spectrogram $M_D \in \mathbb{R}^{N \times L}$ is generated by the element-wise product of $D_T$ and $D_P$ as follows \cite{atk}:
\begin{equation}
    M_D=D_T \odot D_P = MG(T, P)
    \label{eq1}
\end{equation}
where $MG(\cdot)$ indicate mel-generator. 
The mel-generator was trained with $L_1$ and binary divergence loss $L_{bd}$ \cite{dctts, atk} between the ground truth mel-spectrogram $M \in \mathbb{R}^{N \times L}$ and the predicted mel-spectrogram $M_D$. Additionally, we used an auxiliary loss to extract the text and pitch information included in the mel-spectrograms. The auxiliary loss is the sum of $L_1$ losses between $M$ and $D_P$ and between $M$ and $D_T$. We experimentally found that the mel-generator could not model by separating linguistic features and harmonic features without auxiliary loss, and this loss had an effect at the beginning of training. Therefore, we reduced the weight of this auxiliary loss by multiplying the decay rate $\lambda_{init}$ during training. The total loss $L_{mg}$ for the mel-generator can be calculated as follows:
\begin{equation}
    \begin{split}
        &L_{md}(M, M_D) = L_1 \left ( M, M_{D} \right ) + L_{bd} \left ( M, M_{D} \right )\\
        &L_{init}(M, D_P, D_T) = \left (L_1 \left (M, D_P \right ) + L_1\left (M, D_T\right ) \right ) / 2 \\
        &L_{mg} = L_{md} + L_{init} \times \lambda_{init}^{\text{(training stpes)}}\\
    \end{split}
    \label{eq2}
\end{equation}

\subsection{Postnet}
Recent studies have shown that models with a combination of Transformer and CNN modules outperformed models that use them individually \cite{conformer2, conformer}. A Transformer-based on self-attention captures the global context well, but it lacks the ability to extract local features. CNN-based models can efficiently capture local feature patterns, but many layers are needed to obtain global information. Because the adjacent frames are more related in the mel-spectrogram, we designed a CNN-based postnet to capture the local features and generate a more realistic mel-spectrogram from the coarse mel-spectrogram. Figure 2(a) depicts the architecture of the proposed postnet. The postnet of N-Singer, inspired by the ATK's super-resolution network \cite{atk}, consists of a $1 \times 1$ convolutional network, dropout \cite{dropout}, and highway network block that consists of highway networks \cite{highway}, convolutions, and exponential linear units (ELU) \cite{elu}. To reflect the input text and pitch information to the generated mel-spectrogram, the postnet uses encoded phoneme $E_T$ and pitch sequence $E_P$ of the intermediate outputs of encoders as local condition inputs using the local conditioning method as follows \cite{atk}: 
\begin{equation}
\begin{split}
    &C_1 = \phi(U_1*M_D+V_1*E_T+W_1*E_P) \\
    &C_2 = \sigma(U_2*M_D+V_2*E_T+ W_2*E_P) \\
    &M_P = C_1 \odot C_2
\end{split}
\vspace{-0.05cm}
\end{equation}
where, $U_{*}, V_{*}, W_{*}$ are convolutional layers, $\phi$ is ELU activation function and $\sigma$ is sigmoid activation function.
We used $L_1$ and $L_{bd}$ between the ground-truth mel-spectrogram $M$, and generated mel-spectrogram $M_P$. We jointly trained the Transformer-based mel-generator and CNN-based postnet using $L_{mp}$ in (\ref{eq4}) to take advantage of the Transformer and CNN.
\begin{equation}
    L_{mp}(M, M_P) = L_1(M, M_P) + L_{bd}(M, M_P)
    \vspace{-0.05cm}
    \label{eq4}
\end{equation}

\subsection{Discriminator}
Because pronunciation is related to voiced segments in singing voices, it is important that the mel-generator predicts the voiced and unvoiced segments in the mel-spectrogram. The voiced segments are periodic and have clean harmonic components that change slowly; while the unvoiced segments are aperiodic and have noise components that change rapidly. Because the voiced and unvoiced segments' characteristics in singing voice are different, we adopted a voicing-aware conditional GAN with projection discriminators \cite{pwgva}. Our discriminators consisted of 1D and 2D convolutions with residual connections. To capture the harmonic structure in the voiced region of the mel-spectrogram, the first discriminator had residual convolutional layers with a large kernel size to increase the receptive field. Note that the model needs large receptive fields to deal with the long-term temporal dependencies of the harmonic structure in a voiced segment. The second discriminator for the unvoiced region was composed of residual convolutional layers with a small kernel size to capture the noise components. Figure 1(c) shows that to provide conditional information into the model, our discriminator took the inner product between the embedded acoustic feature vector converted by 1D convolutional layers and the intermediate output generated from the input mel-spectrogram. Additionally, a sigmoid activation was placed at the end to prevent extreme increases in output and to train the model stably. We took the V/UV mask from the final output such that the value corresponding to the voiced or unvoiced segment was the output. We use loss functions for adversarial training as follows:  
\begin{equation}
\begin{split}
    L_{dis}={}& \mathbb{E}_{p(\mathbf{x})}[\log (1-D(G(\mathbf{x}), \mathbf{c}))], \\ 
    {}&+\mathbb{E}_{p(\mathbf{y})}[\log D(\mathbf{y}, \mathbf{c})], \forall D\in \left \{ D^\text{v}, D^{\text{uv}}\right \} \\
    L_{adv}={}& \frac{1}{2} \mathbb{E}_{p(\mathbf{x})} \left [ \sum_{D\in \left \{ D^\text{v}, D^{\text{uv}}\right \}}  \log (D(G(\mathbf{x}), \mathbf{c})) \right ] \\
\end{split}
\end{equation}
where $D^v$ and $D^{uv}$ denote the voiced and unvoiced discriminators, $G$ indicate parameters of the generator, $\mathbf{x}$ is the input of generator, $\mathbf{y}$ is a mel-spectrogram from real data, and $\mathbf{c}$ denotes the conditional acoustic feature. Our final training objectives $L_{G}$ and $L_{D}$ for generator and discriminator are follows:
\begin{equation}\label{eq6}
\begin{split}
    &L_{G} = L_{mg} + \lambda_{p} \cdot L_{mp} + \lambda_{adv} \cdot L_{adv} \\
    &L_{D} = \lambda_{adv} \cdot L_{dis} \\
\end{split}    
\end{equation}
where $\lambda_{p}$ and $\lambda_{adv}$ are hyperparameters that need to be adjusted in the experiment.

\section{Experiments}
\label{sec:experiments}

\subsection{Experimental Setup}
\subsubsection{Datasets}

Our singing voice dataset consisted of 50 Korean pop songs and MIDI files collected from a professional female singer. All  recordings were sampled at 48 kHz with 16-bit quantization and then down-sampled to 24 kHz. All songs were split into 1,083 clips, each being 5 to 10 seconds long. We randomly chose 47 songs with 983 clips for training and the remaining three songs with 100 clips for the test. For acoustic features, we extracted F0 sequences and a 1D V/UV (0-1 value) flags from real audio samples using an open-source WORLD vocoder. Mel-spectrograms with 80 bins were extracted by short-time Fourier transformation (STFT) using FFT size, window size, and frame shifts of 2,048, 1,200, and 240, respectively.

\subsubsection{Model Configuration}
The proposed N-Singer system has two encoder and decoder modules. The encoder and decoder consist of six FFT blocks. The dimensions of the phoneme and pitch embeddings, the self-attention's hidden size, and 1D convolution in the FFT block are all set to 256. The number of attention heads is set to 2. The kernel sizes of the two 1D convolution layers in the FFT block are both set to 13, with an input/output size of 256/1024 for the first layer and 1024/256 for the second layer. The output linear layer converts the 256-dimensional output into an 80-dimensional mel-spectrogram. The postnet in N-Singer consisted of a 1D convolutional layer and a highway network block. The highway network block consisted of four highway network layers with hidden size/kernel size set to 256/3, dropout with a rate of 0.05, and a 1D convolutional layer with ELU activation.

The ATK's super-resolution network and discriminator were modified to predict the mel-spectrograms. The output channel of the last convolutional layer of the super-resolution network was set to 80, and the kernel sizes of the two 2D convolutional layers in the discriminator were set to 7$\times$1/3$\times$1.

Parallel WaveGAN with voicing-aware conditional discriminators \cite{pwgva} is used to synthesize the singing voice waveform as a vocoder. Particularly, because the voiced segment length of the singing voice is longer than that of speech, the kernel size of each 1D convolutional layer in the voiced discriminator is set to 13 to have a large receptive field.

\subsubsection{Training}

The three modules of N-Singer, mel-generator, postnet, and discriminator, were trained jointly, while the vocoder was trained separately. The N-Singer model was trained for 100,000 steps with a minibatch size of 8 using the Adam optimizer \cite{adam} with $\beta_1=0.9$, $\beta_2=0.999$, and $\epsilon=10^{-6}$. The initial learning rate was set to $2 \times 10^{-4}$ and was reduced by half for every 50,000 steps. The decay rate in (\ref{eq2}) was set to $0.999$. To warm up the generator, we set the hyperparameters in (\ref{eq6}) as follows \cite{atk}:  
\begin{equation}
\begin{split}
    &\lambda_p = \min(\text{training step}/1000, 1.0)\\
    &\lambda_{adv} = \min(0.01 \times \text{training step}/1000), 1)\\
\end{split}
\end{equation}
The generator of the ATK model was trained for 100,000 steps with a minibatch size of 8 using the Adam optimizer with the same hyperparameters and learning rate decay method as in \cite{atk}. The vocoder was trained for 600,000 steps using a RAdam \cite{radam} optimizer with the same hyperparameters and training method as in \cite{pwgva} on NVIDIA V100 GPU.

\subsection{Evaluation}
In this experiment, we compared the performance of our proposed model N-Singer, quantitatively and qualitatively, with that of the baseline ATK. Some samples are available online\footnote{https://nc-ai.github.io/speech/publications/nsinger/}.

\begin{table}[t]
  \caption{MOS scores}
  \vspace{-0.1cm}
  \label{tab:mos}
  \centering
  \renewcommand{\arraystretch}{0.9}
  \begin{tabular}{cccc}
    \toprule
    \textbf{} & \textbf{Baseline}  & \textbf{N-Singer} & \textbf{GT}  \\
    \midrule
          Pronun.acc & 3.72$\pm$0.16 &  4.23$\pm$0.14 & 4.40$\pm$0.14  \\
        Sound.quality & 3.80$\pm$0.16 &  3.85$\pm$0.16 & 3.85$\pm$0.16  \\
    \bottomrule
  \end{tabular}
\end{table}

\begin{table}[t]
  \caption{Quantitative evaluation results}
  \vspace{-0.1cm}
  \label{tab:Quantitative}
  \centering
  \renewcommand{\arraystretch}{0.9}
  \renewcommand{\tabcolsep}{5.6mm}
  \begin{tabular}{cccc}
    \toprule
    \textbf{Metric} & \textbf{Baseline}  & \textbf{N-Singer}\\
    \midrule
          Accuracy  & 0.886 &  \textbf{0.889}  \\
         Precision  & 0.905 &  \textbf{0.914}  \\
            Recall  & 0.884 &  0.884  \\
          F1-score  & 0.887 &  \textbf{0.893}  \\
    \midrule
          F0 CORR  & 0.984  &  \textbf{0.985}  \\
          F0 RMSE  & 8.959  &  \textbf{8.870}  \\
  V/UV Error (\%) & 10.231 &  \textbf{9.483}   \\
    \bottomrule
  \end{tabular}
  \vspace{-0.3cm}
\end{table}

\subsubsection{Qualitative evaluation}

We conducted a mean opinion score (MOS) listening test as qualitative evaluation. We randomly selected 20 lyrics and MIDI pairs. Thirty native Korean participants evaluated synthesized singing voices based on the criteria for pronunciation accuracy and sound quality. For accurate pronunciation evaluation, participants first listened to the audio samples, looked at the lyrics, and then evaluated the samples. The MOS results, presented in Table 1, show that N-Singer can synthesize a singing voice with a more accurate pronunciation and higher sound quality than ATK. To prove that N-Singer is superior to ATK, we conducted the Wilcoxon \cite{wilcoxon} signed-rank test and win-draw-loss (w-d-l) comparisons. The w-d-l comparisons counted the number of test samples that N-Singer achieved, in terms of better, equal, or worse pronunciation accuracy as compared to ATK. In the Wilcoxon signed-rank test, the p-value was less than 0.05, and the w-d-l records were obtained as 29-1-0. These results indicate that the difference in the MOS results between the two models is statistically significant. Although MOS sound quality is dependent on the performance of the neural vocoder, it shows that the quality of the N-Singer's mel-spectrograms is similar to that of the real mel-spectrograms.

\subsubsection{Quantitative evaluation}

We conducted an evaluation to measure whether the model reflected the input pitch of the musical score to synthesize singing. We extracted F0 sequences from the synthesized samples using the WORLD \cite{world} vocoder. The F0 sequences were converted to note pitch sequences and then compared with the true note pitch sequences (see Table 2 for the results). For all metrics, the N-Singer's value was higher than that of ATK. This result shows that N-Singer synthesizes singing by accurately reflecting the given MIDI note. To further evaluate the synthesized singing voice quality, we calculated the F0 root mean square error (F0 RMSE), F0 correlation (F0 CORR), and voiced/unvoiced error rate (V/UV error) between the ground-truth and the synthesized singing voices. Table 2 shows that N-Singer achieves a higher performance than ATK. It was observed that the V/UV error of N-Singer was significantly less than that of ATK. To compare the V/UV decision accuracy, we plotted the F0 contour generated by the two models in Figure 2. Figure 2 shows that N-Singer better predicts voiced and unvoiced segments of ground-truth than ATK. Because the V/UV decision affects singing intelligibility, the results suggest that our voicing-aware discriminators of N-Singer are effective in improving the pronunciation of the singing synthesized by the mel-generator.

\subsubsection{Ablation studies}
Here, we show that the N-Singer can model text and pitch information individually. In Figure 3, $D_P$ and $D_T$ are generated by Transformer-based encoders from text and pitch information, respectively. $D_T$ consists of non-periodic components with a structure that determines the pronunciation, whereas $D_P$ is composed of a frequency component. The coarse mel-spectrogram $M_D$ is generated by multiplying $D_T$ and $D_P$. $Mask$, $D_M$, and $\hat{M}$ are the intermediate outputs of the ATK. Because ATK uses 1/4-length mel-spectrograms, the intermediate outputs in Figure 3 show pronunciation and pitch information loss. However, because the N-Singer uses the whole mel-spectrogram, it can synthesize a more accurate singing voice.

\begin{figure}[t]
  \centering
    \includegraphics[width=\linewidth]{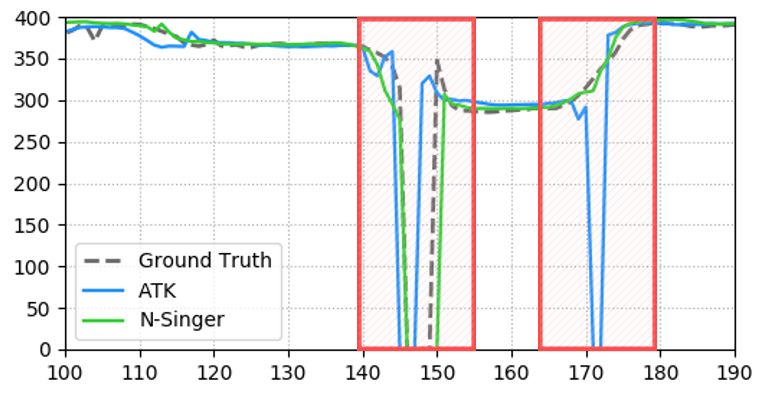}
  \vspace{-0.5cm}
  \caption{F0 contour comparison for N-Singer and ATK}
  \label{fig:f0_contour}
\end{figure}

\begin{figure}[t]
  \centering
  \includegraphics[width=\linewidth]{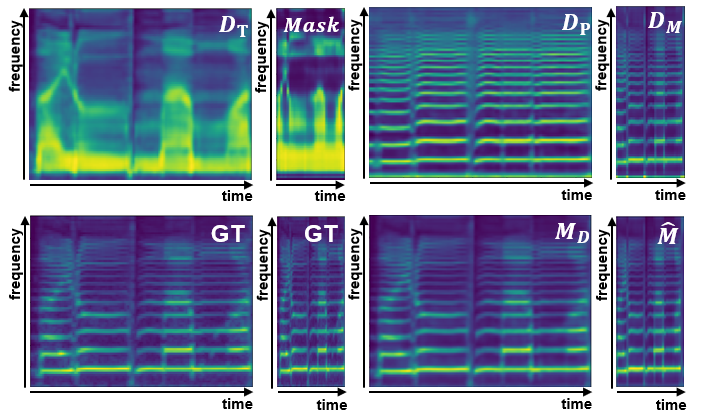}
  \vspace{-0.4cm}
  \caption{Intermediate outputs of N-Singer (left), ATK (right)}
  \label{fig:mel_comparison}
  \vspace{-0.4cm}
\end{figure}

\section{Conclusion}
\label{sec:conclusion}

In this paper, we proposed N-Singer, which can synthesize Korean singing voice with more accurate pronunciation by individually modeling linguistic features of lyrics and musical features of pitches from mel-spectrograms in a non-autoregressive manner.
Experimental results show that N-Singer's voicing-aware discriminators are effective in improving pronunciation.

\vfill\pagebreak

\bibliographystyle{IEEEtran}
\bibliography{refs}

\end{document}